\documentclass[aps,prl,twocolumn,amsmath,superscriptaddress,amssymb,longbibliography]{revtex4}
\usepackage{graphicx}
\usepackage{enumitem}
\usepackage{multirow}
\usepackage[colorlinks,bookmarks=false,citecolor=red,linkcolor=blue,urlcolor=blue]{hyperref}

\usepackage{dcolumn}
\usepackage{bm}

\usepackage{amsmath,amssymb}
\usepackage{tikz,tikz-3dplot}
\usepackage{siunitx}
\DeclareSIUnit{\atmosphere}{atm}
\usepackage{braket}

\usepackage{bbold}

\newcommand{\vu}[1]{\hat{\bm{#1}}}
\newcommand{\etal}{\textit{et al}.}

\newcommand{\plv}[1]{\hat{\mathbf{#1}}}

\newcommand{\bea}{\begin{eqnarray}}          
\newcommand{\eea}{\end{eqnarray}}

\def\r{{\bf r}}

\allowdisplaybreaks

\begin{document}

\title{Metallic bonding in close packed structures: \\structural frustration from a hidden gauge symmetry}
\author{Eric He}
\email{heeric@berkeley.edu}
\affiliation{University of California, Berkeley, California 94720, USA}
\affiliation{Department of Physics, Brock University, St. Catharines, Ontario L2S 3A1, Canada}
\author{C.~M.~Wilson}
\email{cw14mi@brocku.ca}
\affiliation{Department of Physics, Brock University, St. Catharines, Ontario L2S 3A1, Canada}
\author{R. Ganesh}
\email{r.ganesh@brocku.ca}
\affiliation{Department of Physics, Brock University, St. Catharines, Ontario L2S 3A1, Canada}

\date{\today}

\begin{abstract}

Based on its simple valence electron configuration, we may expect lithium to have straightforward physical properties that are easily explained. However, solid lithium, when cooled below 77 K, develops a complex structure that has been debated for decades.
A close parallel is found in sodium below 36 K where the crystal structure still remains unresolved. 
In this letter, we explore a possible driving force behind this complexity. We begin with the observation that Li and Na form close-packed structures at low temperatures. We demonstrate a gauge symmetry that forces \textit{all} close-packed structures to have the same electronic energy and, in fact, the very same band structure. 
This symmetry requires two conditions: (a) bands must arise from $s$ orbitals, and (b) hoppings beyond second-nearest neighbours must be negligible. We argue that both can be reasonably invoked in Li and Na. 
When these conditions are satisfied, we have extensive degeneracy with the number of competing iso-energetic structures growing exponentially with linear system size. 
Weak effects, such as $p$-orbital admixture, long-range hopping and phonon zero-point energy, can break this symmetry. These can play a decisive role in `selecting' one particular ordered structure. 
This point of view may explain the occurrence of ordered structures in Li and Na under pressure. 
Our results suggest that martensitic transitions may also occur in heavier alkali metals such as potassium.
\end{abstract}

\maketitle

{\color{blue} \it{Introduction}}---~
Systems with extensive degeneracy are fertile ground for interesting physical properties. This is best seen in the field of frustrated magnetism where a large number of classical magnetic orders compete\cite{Chalker2011,Ramirez1994}.  
A decisive role is then played by otherwise-small effects such as spin wave entropy, magnon zero-point energy, impurities, etc.\cite{Villain1980,Henley1989,Shender1996} They may `select' one particular ordered state or even give rise to liquid-like disordered states\cite{Balents2010}. In this letter, we bring out an analogue wherein a large number of crystal structures compete with one another. The competition is among close-packed structures, a family of structures with a long history dating from Kepler's conjecture in 1611\cite{Kepler_1611,Szpiro2003kepler}. Their realizations include more than half of all elemental solids\cite{Steurer2001,Arblaster2018}
. We argue that they compete and give rise to structural frustration in lithium and sodium.

The low-temperature, ambient-pressure structure of lithium has been debated for decades. 
At room temperature and pressure, lithium crystallizes in the bcc structure. A martensitic transition occurs when cooled to $77~\si{\kelvin}$, first reported by Barrett in 1947\cite{barrett_PR_1947}. Neutron diffraction measurements of McCarthy~\etal~in 1980 revealed that the new phase is neither fcc nor hcp~\cite{mccarthy_PRB_1980}. In 1984, Overhauser proposed the $9R$ structure, a close-packed structure with a 9-layer unit cell~\cite{overhauser_PRL_1984}. Subsequent neutron experiments found evidence of stacking faults as well as the coexistence of the fcc, hcp, and $9R$ structures at short ranges~\cite{berliner_PhysicaB_1986,smith_PRL_1987,berliner_PRB_1989,smith_PhysicaB_1989,schwarz_PRL_1990}. 
More recently, in 2017, Elatresh~\etal~\cite{elatresh_PNAS_2017} argued that de Haas-van Alphen measurements are inconsistent with the~$9R$ structure. In the same year, a comprehensive study by Ackland~\etal~\cite{ackland_Science_2017} concluded that the actual ground state of Li is fcc. 
A similar picture emerges in sodium, where Barrett demonstrated a martensitic transition below $36~\si{\kelvin}$\cite{Barrett1955,Barrett1956}. Despite multiple studies stretching over decades\cite{Martin1958,Hull1959,Szente1988,berliner_PRB_1989,Berliner1992,Elatresh2020}, the structure at lower temperatures remains unresolved.

These studies reveal competition among close-packed structures in Li and Na. This is surprising given that phases such as fcc, hcp and 9R have entirely different symmetries. There is no \textit{a priori} reason for them to be close in energy. We propose an explanation in the form of a hidden gauge symmetry, building upon little known results of Thorpe\cite{Thorpe1972} and Betteridge\cite{Betteridge1981} that point to degeneracies in tight binding bands of close-packed solids.

{\color{blue} \it{Close-packed structures and their representations}}---~
Close-packed structures are the densest possible arrangements of spheres in three dimensions\cite{KrishnaPandey1981,Conway2013sphere}, with a packing fraction of $\sim$74\%. They are constructed by stacking layers of triangular-lattice arrangements. Each layer must be laterally displaced in one of two directions relative to the layer below.  
This leads to three possible lateral positions for each layer, denoted by $A,B,C$. A close-packed structure can be denoted as a Barlow stacking sequence 
-- a sequence of letters where no two adjacent letters can be the same. For example, fcc is represented as $(ABC)$, a three-letter pattern that repeats indefinitely. 
The number of close-packed structures grows exponentially with the number of layers. As each layer can take one of two letters (to be distinct from the previous layer), $M$ layers can be stacked in $2^{M-1}$ ways.

The H\"agg code\cite{Hagg1943} is a dual representation that maps each close-packed structure onto a 1D Ising configuration. The Ising variable represents a certain `chirality' -- the change of lateral position upon moving along the stacking direction. We define the chirality for $A\rightarrow B\rightarrow C\rightarrow A$ as $+1$, and for $A\rightarrow C\rightarrow B\rightarrow A$ as $-1$. The fcc structure maps to an Ising ferromagnet, with all chirality variables being $+1$ (or all being $-1$). Below, we will consider a structure with a repeating $M$-layer pattern. In the H\"agg code, this yields a sequence of $M$ Ising variables with periodic boundaries: $\sigma_j$, where $j=1,\ldots,M$ with $\sigma_{M+1}\equiv \sigma_1$. The `net chirality', defined as $\sum_{j=1}^M \sigma_j$, will play a crucial role. We will then discuss the general case with no periodicity in stacking.

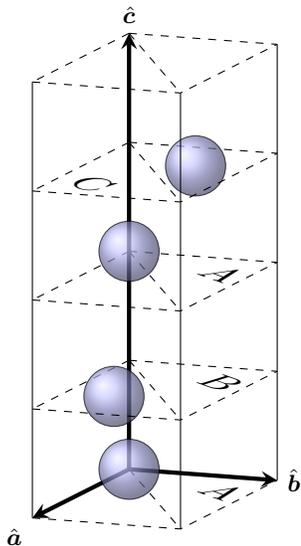
\begin{figure}
    \centering
    \tdplotsetmaincoords{65}{130}     
\begin{tikzpicture}[scale=2,tdplot_main_coords]
  \def\r{0.2}
  
  \draw [canvas is xy plane at z=0,>=stealth,->,ultra thick] (0,0) -- (1,0) node [below left] {$\hat{\bm{a}}$};
  \draw [canvas is xy plane at z=0,>=stealth,->,ultra thick] (0,0) -- (120:1) node [right] {$\hat{\bm{b}}$};
  \draw [>=stealth,->,ultra thick] (0,0,0) -- (0,0,3.2) node [above] {$\hat{\bm{c}}$};

  \draw [canvas is xy plane at z=0,dashed] (0,0) --++ (60:1);
  \foreach \z in {0.8,1.6,2.4,3.2} {
     \draw [canvas is xy plane at z=\z,dashed] (0,0) --++ (0:1);
     \draw [canvas is xy plane at z=\z,dashed] (0,0) --++ (120:1);
     \draw [canvas is xy plane at z=\z,dashed] (0,0) --++ (60:1);
  }

  \node [canvas is xy plane at z=0,rotate=90,transform shape] at (100:0.7) {$A$};
  \node [canvas is xy plane at z=0.8,rotate=90,transform shape] at (100:0.7) {$B$};
  \node [canvas is xy plane at z=1.6,rotate=90,transform shape] at (100:0.7) {$A$};
  \node [canvas is xy plane at z=2.4,rotate=90,transform shape] at (20:0.7) {$C$};

  \tdplottransformmainscreen{0}{0}{0};
  \shadedraw[tdplot_screen_coords, ball color = blue!30, opacity=0.7] (\tdplotresx,\tdplotresy) circle (\r);
  \tdplottransformmainscreen{cos(30)/sqrt(3)}{sin(30)/sqrt(3)}{0.8};
  \shadedraw[tdplot_screen_coords, ball color = blue!30, opacity=0.7] (\tdplotresx,\tdplotresy) circle (\r);
  \tdplottransformmainscreen{0}{0}{1.6};
  \shadedraw[tdplot_screen_coords, ball color = blue!30, opacity=0.7] (\tdplotresx,\tdplotresy) circle (\r);
  \tdplottransformmainscreen{cos(90)/sqrt(3)}{sin(90)/sqrt(3)}{2.4};
  \shadedraw[tdplot_screen_coords, ball color = blue!30, opacity=0.7] (\tdplotresx,\tdplotresy) circle (\r);

  \foreach \z in {0,0.8,1.6,2.4,3.2} {
     \draw [canvas is xy plane at z=\z,dashed] (1,0) --++ (120:1) --++ (180:1);
  }

  \draw (1,0,0) -- (1,0,3.2);
  \draw ({cos(60)},{sin(60)},0) -- ({cos(60)},{sin(60)},3.2);
  \draw ({cos(120)},{sin(120)},0) -- ({cos(120)},{sin(120)},3.2);
\end{tikzpicture}
    \caption{Hexagonal unit cell of $ABAC$ close-packing. There are four basis atoms, one from each layer. The basis atom from layer $B$ ($C$) is located at the centroid of the `up'-triangle (`down'-triangle).}
    \label{fig:unit-cell}
\end{figure}

{\color{blue} \it{Geometry of close-packing}}---~
We consider a stacking sequence which repeats periodically along the $\plv{z}$ axis after~$M$ layers. The underlying Bravais lattice is hexagonal, with primitive lattice vectors 
\begin{equation}
    \plv{a}=\plv{x},\quad
    \plv{b}=R\!\left(\tfrac{2\pi}{3}\right)\plv{a},\quad
    \plv{c}=M\sqrt{\frac23}\,\plv{z},
    \label{eq:primitive-vectors}
\end{equation}
where $R(\theta)$ denotes counterclockwise rotation about $\bm{\hat{c}}$ by angle~$\theta$. The spacing between adjacent layers in any closed-packed solid is $\sqrt{\frac23}(2r)$, where $r$ is the atomic radius. The units of length in Eq.~(\ref{eq:primitive-vectors}) are chosen such that~$2r=1$. The unit cell contains $M$ basis atoms, one from each layer, with an example shown in Fig.~\ref{fig:unit-cell}.

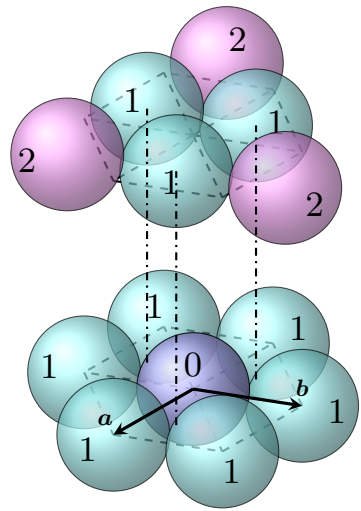
\begin{figure}
    \centering
    \tdplotsetmaincoords{55}{135}     
\begin{tikzpicture}[tdplot_main_coords,scale=1.5]
  \def\d{2.75}
  \def\f{1.32}


  \tikzset{A/.pic={
  \foreach \t in {0,60,...,300} {
    \draw [dashed,thick] ({cos(\t)},{sin(\t)},0) -- ({cos(\t+60)},{sin(\t+60)},0);
    \draw [dashed,thick] (0,0,0) -- ({cos(\t-180)},{sin(\t-180)},0);
    }
  \foreach \t in {240,180,300,120} {
    \tdplottransformmainscreen{cos(\t)}{sin(\t)}{0}
    \shadedraw[tdplot_screen_coords, ball color = cyan!40,opacity=0.7] (\tdplotresx,\tdplotresy) circle (0.5) node [opacity=1] at ({\f*\tdplotresx},{\f*\tdplotresy)},0) {\small 1};
    }
  \tdplottransformmainscreen{0}{0}{0}
  \shadedraw[tdplot_screen_coords, ball color = blue!30,opacity=0.9] (\tdplotresx,\tdplotresy) circle (0.5) node [opacity=1,above] at (0,0) {\small 0};
  \foreach \t in {0,60} {
    \tdplottransformmainscreen{cos(\t)}{sin(\t)}{0}
    \shadedraw[tdplot_screen_coords, ball color = cyan!40,opacity=0.7] (\tdplotresx,\tdplotresy) circle (0.5) node [opacity=1] at ({\f*\tdplotresx},{\f*\tdplotresy)},0) {\small 1};
    }
  }}
  \tikzset{B/.pic={
  \foreach \t in {0,60,...,300} {
    \draw [dashed,thick] ({cos(\t)},{sin(\t)},0) -- ({cos(\t+60)},{sin(\t+60)},0);
    \draw [dashed,thick] (0,0,0) -- ({cos(\t-180)},{sin(\t-180)},0);
    }
  \foreach \t in {270} {
    \tdplottransformmainscreen{2/sqrt(3)*cos(\t)}{2/sqrt(3)*sin(\t)}{0}
    \shadedraw[tdplot_screen_coords, ball color = magenta!40,opacity=0.7] (\tdplotresx,\tdplotresy) circle (0.5) node [opacity=1]  at ({\f*\tdplotresx},{\f*\tdplotresy)},0) {\small 2};
    }
  \foreach \t in {210,330,90} {
    \tdplottransformmainscreen{1/sqrt(3)*cos(\t)}{1/sqrt(3)*sin(\t)}{0}
    \shadedraw[tdplot_screen_coords, ball color = cyan!40,opacity=0.7] (\tdplotresx,\tdplotresy) circle (0.5) node [opacity=1] at ({\f*\tdplotresx},{\f*\tdplotresy)},0) {\small 1};
    }
  \foreach \t in {30,150} {
    \tdplottransformmainscreen{2/sqrt(3)*cos(\t)}{2/sqrt(3)*sin(\t)}{0}
    \shadedraw[tdplot_screen_coords, ball color = magenta!40,opacity=0.7] (\tdplotresx,\tdplotresy) circle (0.5) node [opacity=1] at ({\f*\tdplotresx},{\f*\tdplotresy)},0) {\small 2};
    }
  }}
  \tikzset{C/.pic={
  \foreach \t in {0,60,...,300} {
    \draw [dashed,thick] ({cos(\t)},{sin(\t)},0) -- ({cos(\t+60)},{sin(\t+60)},0);
    \draw [dashed,thick] (0,0,0) -- ({cos(\t-180)},{sin(\t-180)},0);
    }
  \foreach \t in {210} {
    \tdplottransformmainscreen{2/sqrt(3)*cos(\t)}{2/sqrt(3)*sin(\t)}{0}
    \shadedraw[tdplot_screen_coords, ball color = magenta!40,opacity=0.7] (\tdplotresx,\tdplotresy) circle (0.5) node [opacity=1] at ({\f*\tdplotresx},{\f*\tdplotresy)},0) {\small 2};
    }
  \foreach \t in {150,270,30} {
    \tdplottransformmainscreen{1/sqrt(3)*cos(\t)}{1/sqrt(3)*sin(\t)}{0}
    \shadedraw[tdplot_screen_coords, ball color = cyan!40,opacity=0.7] (\tdplotresx,\tdplotresy) circle (0.5) node [opacity=1] at ({\f*\tdplotresx},{\f*\tdplotresy)},0) {\small 1};
    }
  \foreach \t in {330,90} {
    \tdplottransformmainscreen{2/sqrt(3)*cos(\t)}{2/sqrt(3)*sin(\t)}{0}
    \shadedraw[tdplot_screen_coords, ball color = magenta!40,opacity=0.7] (\tdplotresx,\tdplotresy) circle (0.5) node [opacity=1] at ({\f*\tdplotresx},{\f*\tdplotresy)},0) {\small 2};
    }
  }}

\path [transform shape] (0,0,0) pic {A};
\path [transform shape] (0,0,{\d}) pic {C};

\foreach \t in {30,150,270} {\draw [thick,dashdotted] ({1/sqrt(3)*cos(\t)},{1/sqrt(3)*sin(\t)},0) --++ (0,0,\d);};

\draw [very thick,>=stealth,->] (0,0,0) -- (1,0,0) node [above,xshift=-3pt] {$\bm{a}$};
\draw [very thick,>=stealth,->] (0,0,0) -- ({cos(120)},{sin(120)},0) node [above] {$\bm{b}$};
\end{tikzpicture}
    \caption{Local atomic environment in the sequence $\dots AB\dots$. Layers are offset in the stacking direction for clarity. The reference atom is labelled `$0$'. Its nearest and second-nearest neighbours are labelled `$1$' and `$2$' respectively. }
    \label{fig:local-env}
\end{figure}

In any close-packed structure, all atoms have a similar local environment. There are twelve nearest-neighbours (1nn), that can be divided into three sets: 
(i) Within the same layer, there are six 1nn located at corners of a hexagon centred at the reference atom as shown in Fig.~\ref{fig:local-env}. For later use, we denote the vectors connecting the centre of the hexagon to its corners as $\mathcal{N}_\parallel$. (ii) Three 1nn are located in the layer above, shown in Fig.~\ref{fig:local-env}. Their relative positions belong to one of two possible sets, depending on the chirality between the two layers. We denote these sets as $\mathcal{N}_{\perp}^+$ and $\mathcal{N}_{\perp}^-$. Each set contains three vectors, with all vectors having the same projection along the stacking direction (same $\hat{z}$-component). (iii) Finally, there are three 1nn in the layer below.  

There are six second-nearest neighbours (2nn): three in the layer above (see Fig.~\ref{fig:local-env}) and three in the layer below. Their positions relative to the reference atom depend on the chirality between the respective layers. We define two sets of 2nn vectors to the neighbours in the layer above: $\mathcal{N}_{2,\perp}^+$ and $\mathcal{N}_{2,\perp}^-$.

For third (3nn) and further neighbours, the environment may differ from one close-packed structure to another. For example, in an hcp structure, there are two 3nn located at precisely $\pm 2\sqrt{\frac23} \hat{z}$. In fcc, there are no atoms located at these relative positions. The supplementary material compiles this information in the form of a neighbour table\cite{supplementary}.

{\color{blue} \it{Tight binding description}}---~
In fcc structure, Li and Na have nearly spherical Fermi surfaces arising from a single band with dominant $s$-orbital character\cite{elatresh_PNAS_2017,Elatresh2020}. In a generic close-packed structure, we describe bands arising from a single $s$ orbital in each atom's valence shell. For simplicity, we proceed assuming that orbitals on distinct atoms are orthogonal to one another. If this assumption is relaxed, the symmetry described below will still hold provided only 1nn bonds are retained. A detailed discussion is presented in the supplementary material\cite{supplementary}.

Each orbital can be written as $\vert u, v, w,\alpha\rangle$, where $(u,v,w)$ are integers that pick a unit cell.  
The index $\alpha\in[1,M]$ picks one atom from the $M$-atom basis. 
Assuming electrons can hop to 1nn and 2nn, the Hamiltonian acts
as follows:
\begin{eqnarray}
\nonumber    \hat{H}\ket{u,v,w,\alpha}
    =&-&t\sum_{(u_1,v_1,w_1,\alpha_1) \in 1nn} \ket{u_1,v_1,w_1,\alpha_1}\\
    &-&t'\sum_{(u_2,v_2,w_2,\alpha_2) \in 2nn} \ket{u_2,v_2,w_2,\alpha_2},~~~~~
    \label{eq:tb_hamiltonian}
\end{eqnarray}
where~$t$ and $t'$ are hopping amplitudes to 1nn and 2nn respectively. The accompanying summations are over nearest and next-nearest neighbours of $(u,v,w,\alpha)$. 
We propose the following Bloch wave ansatz for the stationary states of~$\hat{H}$:
\begin{equation}
    \ket{\bm{k}}   \propto\sum_{u,v=1}^L\sum_{w=1}^{L_c}\sum_{\alpha=1}^M c_\alpha\,e^{i\bm{k}\cdot\bm{R}_{uvw\alpha}}\ket{u,v,w,\alpha},
    \label{eq:bloch-ansatz}
\end{equation}
where $c_\alpha$ are expansion coefficients to be determined, and $\bm{R}_{uvw\alpha}$ is the position vector of the atom $(u,v,w,\alpha)$. 
Asserting that Eq.~\ref{eq:bloch-ansatz} is an eigenstate of $\hat{H}$, we are led to diagonalizing the following matrix,
\begin{equation}
    H(\bm k)=
    \begin{pmatrix}
    V_D & V_1 & 0 & \cdots & V_M^*\\
    V_1^* & V_D & V_2 & \ddots & 0\\
    0 & \ddots & \ddots & \ddots & 0\\
    \vdots & & V_{M-2}^* & V_D & V_{M-1} \\
    V_M & 0 & \cdots & V_{M-1}^* & V_D
    \end{pmatrix},
    \label{eq:Bhamiltonian}
\end{equation}
where each entry is a function of $\bm k$ (argument suppressed for brevity). The diagonal entries are identical, given by $V_D    =V_0-t\sum_{\bm{\eta}\in \mathcal{N}_\parallel}e^{i\bm{k}\cdot\vu{\eta}}$, where $V_0$ is the on-site contribution and the remaining terms arise from hopping to 1nn within the same layer. 
The off-diagonal entries depend on chirality variables. Denoting the close-packed structure as $(\sigma_1,\ldots,\sigma_M)$ in H\"agg code notation, the right-of-diagonal entry $V_j$ depends on $\sigma_j$. It takes two possible values given by
\begin{align}
\nonumber    
    V_+
    =-t\sum_{\bm{\eta}\in \mathcal{N}_\perp^+}e^{i\bm{k}\cdot\vu{\eta}} -t'\sum_{\bm{\eta}\in \mathcal{N}_{2,\perp}^+}e^{i\bm{k}\cdot\vu{\eta}};\\
    V_-
    =-t\sum_{\bm{\eta}\in \mathcal{N}_\perp^-}e^{i\bm{k}\cdot\vu{\eta}}
    -t'\sum_{\bm{\eta}\in \mathcal{N}_{2,\perp}^-}e^{i\bm{k}\cdot\vu{\eta}}.
    \label{eq.Vs}
\end{align}
These terms encode hoppings to the next layer, including 1nn and 2nn. Crucially, $V_+$ and $V_-$ have the same amplitude but may differ in phase, see Supplement\cite{supplementary}. For later use, we write 
\begin{eqnarray}
V_+ (\bm k) = V_-(\bm k)e^{2i\phi(\bm k)},
\label{eq.defphi}
\end{eqnarray} 
where $\phi(\bm k)$ is a $\bm k$-dependent phase. The eigenvalues of Eq.~(\ref{eq:Bhamiltonian}), evaluated at each $\bm k$ within the Brillouin zone, give the electronic band structure.

{\color{blue} \it{Gauge symmetry with periodic boundaries}}---~
Consider two distinct close-packed structures denoted as $(\sigma_1,\ldots,\sigma_M)$ and $(\chi_1,\ldots,\chi_M)$ in H\"agg code notation. As both have the same number of layers, they share the same primitive lattice vectors as given in Eq.~(\ref{eq:primitive-vectors}) and consequently, the same Brillouin zone. For a given momentum within this Brillouin zone, $\bm k$, we define their Bloch Hamiltonians as $H_\sigma (\bm k)$ and $H_\chi (\bm k)$, both of the form given in Eq.~(\ref{eq:Bhamiltonian}). Remarkably, 
these Hamiltonians are related by a unitary gauge transformation \textit{if} the two configurations have the same net chirality, i.e., if $\sum_j \sigma_j = \sum_j \chi_j$. To demonstrate this, we propose a transformation matrix
\begin{equation}
    W_{\sigma,\chi}=\mbox{diag}\left(e^{i\theta_1},e^{i\theta_2},\dots,e^{i\theta_M}\right),
    \label{eq.W}
\end{equation}
where $\{\theta_j\}$ are to be determined so as to satisfy 
\begin{eqnarray}
W_{\sigma,\chi}^\dagger H_{\sigma}(\bm k) W_{\sigma,\chi} = H_\chi(\bm k).
\label{eq.transform}
\end{eqnarray}
From the form of the generic Hamiltonian in Eq.~(\ref{eq:Bhamiltonian}), we see that the diagonal terms ($V_D$'s) are preserved under the transformation by $W$. These terms are identical in $H_{\sigma}(\bm k) $ and $H_\chi(\bm k)$. We then examine off-diagonal terms. We denote the right-of-diagonal terms in these two Hamiltonians as $V_j (\sigma_j)$ and $V_j (\chi_j)$ respectively. If $\sigma_j = \chi_j$, they are the same. If not, one must be given by $V_+$ and the other by $V_-$, both defined in Eq.~(\ref{eq.Vs}). 
 In order to satisfy Eq.~(\ref{eq.transform}), we must have $V_j (\sigma_j)\,e^{i\left(\theta_{j+1}-\theta_{j}\right)}= V_j(\chi_j)$. This reduces to 
\begin{equation}
(\theta_{j+1}-\theta_{j}) = ( \chi_j-\sigma_j)~\phi(\bm k),
\label{eq.delta_theta}
\end{equation}
where $\phi(\bm k)$ is defined in Eq.~\ref{eq.defphi}. These relations determine $\theta_j$'s. If Eq.~(\ref{eq.delta_theta}) is summed over $j$'s, the left-side vanishes due to telescopic cancellation. In order to have consistent $\theta_j$ values, the right side must also vanish when summed. This can only happen when $\sum_j \sigma_j = \sum_j \chi_j$. We have arrived at a constraint for when $W_{\sigma,\chi}$ can be chosen to satisfy Eq.~(\ref{eq.transform}):
the two close-packed structures must have the same net chirality.

For two close-packed structures that are periodic with the same number of layers and have the same net chirality, the Bloch Hamiltonians at each momentum are unitary-related. It follows that the two structures have precisely the same band structure. When these bands are filled by placing the system in contact with an electron reservoir, the two structures will have the same electronic energy for any chemical potential and temperature. This situation is reminiscent of Ising magnets where the energy only depends on the net magnetization\cite{Kittel}. 
The configuration space separates into sectors characterized by net chirality, some having large degeneracies. An example with $M=18$ layers is discussed in the supplementary material\cite{supplementary}.

{\color{blue} \it{Gauge symmetry with open boundaries}}---~
We now consider open, rather than periodic, boundary conditions along the stacking direction. This choice is more realistic and applies to single-grain crystals grown in experiments.

We may use the Hamiltonian matrix of Eq.~\ref{eq:Bhamiltonian}, but with the corner entries, $V_M$ and $V_M^*$, set to zero. In this setting, \textit{all} close-packed structures are degenerate regardless of chirality. Given any pair of structures, we can always construct a transformation matrix of the form in Eq.~(\ref{eq.W}). There are $M-1$ relations of the form Eq.~(\ref{eq.delta_theta}) that constrain $M$ variables ($\theta_j$'s) -- a non-trivial solution always exists.  

The symmetry can also be seen from an alternative argument. At each $\bm k$, the band energies are roots of the characteristic equation $\det\big\{H(\bm k)-E\big\}=0$. Since $H(\bm k)$ is tridiagonal, the determinant can be evaluated using transfer matrices~\cite{molinari_LinAlgApp_2008}:
\begin{equation*}
    \det(H-E)
    =\left[
    \prod_{j=2}^{M}
    \begin{pmatrix}
        V_D-E & -|V_j|^2 \\ 1 & 0
    \end{pmatrix}
    \begin{pmatrix}
        V_D-E & 0 \\ 1 & 0
    \end{pmatrix}
    \right]_{11}.
\end{equation*}
As we have already established, $|V_+|^2=|V_-|^2$. There is nothing in the matrix product above which depends on the particular stacking sequence. For all $M$-layer sequences, the eigenenergies are the roots of the same characteristic equation. If these levels are filled, all stacking sequences will have the same energy irrespective of the chemical potential and temperature. We conclude that all close-packed structures with the same number of layers and open ends are degenerate.

{\color{blue} \it{Symmetry breaking effects}}---~We have demonstrated a large ground state degeneracy that emerges from a gauge symmetry. Our arguments are based on standard assumptions in any discussion of band structure, viz., translational symmetry (within layers), the Born-Oppenheimer approximation, negligible interactions, etc. 
We now explore some effects that could disrupt the gauge symmetry and lead to ordering.

(i) Higher orbitals:
Our arguments are based on a tight binding description of $s$ orbitals. With higher orbitals such as $p$, $d$, etc., the tridiagonal form of Eq.~\ref{eq:Bhamiltonian} will still hold (assuming hopping is truncated at second nearest-neighbours). However, the elements of Eq.~\ref{eq:Bhamiltonian} will be promoted to matrices, e.g., if each atom contributes one $s$ and 3 $p$ orbitals, 
we will have $4\times 4$ blocks with entries that can be written systematically using Slater-Koster parameters\cite{slater_PR_1954}. With multiple entries in each block, a simple gauge transformation of the form Eq.~(\ref{eq.W}) can no longer relate two different Bloch Hamiltonians. 

Surprisingly, numerical results indicate that an approximate degeneracy survives when $p$ orbitals are included. In the supplementary material, we discuss the band structure of lithium with $2s$ and $2p$ orbitals included. Band structures for distinct close-packed structures are very close to one another, particularly when the net chirality is the same.

(ii) Phonon zero point energies: Our arguments are based solely on electronic energy.  
Standard calculations of phonon band structures do not show any degeneracy among close-packed structures, see supplement\cite{supplementary}. We consider the $T\rightarrow 0$ limit, where phonons can only contribute to energy via a zero-point contribution. We estimate this contribution to be a few percent of the electronic energy\cite{supplementary}. Nevertheless, it may break the degeneracy and select a certain close-packed structure. This could explain the recent observation of fcc order in lithium at low-temperatures\cite{ackland_Science_2017}. In frustrated magnets, it is well known that zero-point energies of spin waves can lead to `state selection'\cite{Rau2018,Khatua2021}. With structural frustration, phonon zero-point energies play an analogous role. 

At finite temperatures, phonons bands are occupied according to the Bose-Einstein distribution. They contribute to energy and entropy. Their free energy may break the degeneracy or even favour an entirely different structure, e.g., see Ref.~\cite{Woodcock1997}. This could explain the appearance of non-close-packed bcc order above a critical temperature\cite{Martin1959}.

(iii) Long-ranged hopping: The gauge symmetry only applies if hopping beyond second neighbours is negligible. If the solid is pressurized, atoms will move closer and longer-range processes will emerge. This may explain the fact that Li and Na order under pressure\cite{Vaks1989,Schaeffer2015}.
To estimate the relevance of long-ranged hopping under ambient pressure, we evaluate a ratio, $\rho = a/a_0$, where $a$ is the experimentally determined lattice constant and $a_0$ is the atomic radius. A large value, $\rho \gg 1$, indicates that atoms are well-separated and long-ranged hoppings are negligible. We find $\rho_{Li}\sim 2.31$ and $\rho_{Na} \sim2.26$. As they are greater than unity, it is reasonable to neglect long-ranged hoppings. In fact, heavier alkali metals also have similar values with $\rho_{K}~\sim 2.31$, $\rho_{Rb}~\sim 2.24$ and $\rho_{Cs}~\sim 2.14$. This suggests that they may also show frustration and a martensitic transition, perhaps requiring longer annealing times than explored previously.

{\color{blue} \it{Discussion}}---~Upon cooling, solids typically become more ordered in their structure. In systems that undergo a disordering transition\cite{Nix1938,Kozlov1976}, the ordered phase is at low temperatures while the disordered phase appears at high temperatures. Lithium and sodium are two striking counterexamples. They lose order when cooled, with the disordered phase seemingly extending to zero temperature. 
We provide an explanation in terms of a gauge symmetry that connects an infinite family of idealized close-packed structures.

Our arguments based on a tight binding approach can be compared with the substantive ab initio literature on ordering in lithium and sodium\cite{maysenholder_PRB_1985,dacorogna_PRB_1986,Sankaran_1992,Hutcheon2019,Raju2019}. 
Unlike ab initio studies which are restricted to small system sizes, our tight binding approach treats all close-packed structures on the same footing. 
Boundary conditions play an important role -- periodic boundaries along the stacking direction artificially limit the degeneracy. 
An interesting future direction is to examine whether the small energy differences seen in ab initio studies are affected by choice of boundary conditions.

{\color{blue} \it{Acknowledgments}.}---~
We thank K. Samokhin for helpful discussions. This work was supported by the Natural Sciences and Engineering Research Council of Canada through Discovery Grant 2022-05240.

\bibliographystyle{apsrev4-1} 
\bibliography{refs}

\newpage \clearpage
 
\onecolumngrid 
\setcounter{secnumdepth}{3}

\renewcommand{\theequation}{S\arabic{equation}}
\renewcommand{\thefigure}{S\arabic{figure}}
\begin{center}
 \textbf{\large Supplemental Material for ``Metallic bonding in close packed structures: \\structural frustration from a hidden gauge symmetry''}\\[.5cm]
E. He,$^1,2$ C. M. Wilson$^1$ and R. Ganesh$^1$\\[.4cm]
{\itshape ${}^1$Department of Physics, Brock University, St. Catharines, Ontario L2S 3A1, Canada\\
\itshape ${}^2$University of California, Berkeley, California 94720, USA
}
(Dated: \today)\\[2cm]
\end{center}

\section{Neighbours in close-packed arrangements}
In the main text, we have described the arrangement of nearest and next-nearest neighbours in any close-packed arrangement. We compile this information in the form  of a table in Tab.~\ref{tab.neighbours1}.

\begin{table}[h!]
\centering
\begin{tabular}{|c|c|c|c|c|}
\hline
Neighbour & Distance & Layer  & Multiplicity & Representative position vector\\
\hline
\hline
\multirow[c]{3}{*}{1nn} & \multirow[c]{3}{*}{1} & $0$ & $6$ & $(1,0,0)$ \\
& & $+1$ & $3$ & $(\frac{1}{2},\frac{1}{2\sqrt{3}},\sqrt{\frac{2}{3}})$\\
& & $-1$ & $3$ & $(\frac{1}{2},\frac{1}{2\sqrt{3}},-\sqrt{\frac{2}{3}})$\\\hline
\multirow[c]{2}{*}{2nn} & \multirow[c]{2}{*}{$\sqrt{2}\sim 1.41$} & $+1$ & $3$ & $(0,\frac{2}{\sqrt{3}},\sqrt{\frac{2}{3}})$\\
& & $-1$ & $3$ & $(0,\frac{2}{\sqrt{3}},-\sqrt{\frac{2}{3}})$\\ \hline
\end{tabular}
\caption{Nearest (1nn) and next-nearest neighbours (2nn) in any close-packed structure. `Distance' represents distance from the reference atom to the neighbour in question. `Layer' represents the layer where the neighbour resides, with the reference atom taken to be in layer 0. `Multiplicity' indicates the number of neighbours located at that distance in that layer. The last column gives the relative position for a representative neighbour. 
}
\label{tab.neighbours1}
\end{table}

When identifying third and further neighbours, the hierarchy depends on the structure in question. This can be seen from Tab.~\ref{tab.neighbours2} which tabulates neighbours beyond those listed in Tab.~\ref{tab.neighbours1}. For example, in the hcp structure with repeated $ABA$ patterns, every atom has its third neighbour located at a distance of $2\sqrt{\frac{2}{3}} \sim 1.63$ units. In an fcc structure with no $ABA$ patterns, third neighbours are at a distance of $\sqrt{3}\sim 1.73$ units. In a generic structure with some $ABA$ and some $ABC$ patterns, some atoms will have a third neighbour at $\sim 1.63$ units while the others will have third neighbours at $\sim 1.73$ units.

\begin{table}[h!]
\centering
\begin{tabular}{|c|c|c|c|c|}
\hline
Local structure & Distance & Layer  & Multiplicity & Representative position vector\\
\hline\hline
\underline{A}BA  & $2\sqrt{\frac{2}{3}}\sim 1.63$ & $+2$    & 1 & 
(0,0,$2\sqrt{\frac{2}{3}}$)   \\
\hline
AB\underline{A}  & $2\sqrt{\frac{2}{3}}\sim 1.63$ & $-2$    & 1 & 
(0,0,-$2\sqrt{\frac{2}{3}}$)   \\
\hline
All structures & $\sqrt{3}\sim 1.73$ & 0 & 6 & ($\frac{3}{2}$,$\frac{\sqrt{3}}{2}$,0)\\
\hline
All structures& $\sqrt{3}\sim 1.73$ & $+1$ & 6 & ($\frac{3}{2}$,$\frac{1}{2\sqrt{3}}$,$\sqrt{\frac{2}{3}}$)\\
\hline
All structures & $\sqrt{3}\sim 1.73$ & $-1$ & 6 & ($\frac{3}{2}$,$\frac{1}{2\sqrt{3}}$,$-\sqrt{\frac{2}{3}}$)\\
\hline
\underline{A}BC  & $\sqrt{3}\sim 1.73$ & $+2$ & 3 & 
($\frac{1}{2}$,$\frac{-1}{2\sqrt{3}}$,$2\sqrt{\frac{2}{3}}$)   \\
\hline
CB\underline{A}  & $\sqrt{3}\sim 1.73$ & $-2$ & 3 & 
($\frac{1}{2}$,$\frac{-1}{2\sqrt{3}}$,$-2\sqrt{\frac{2}{3}}$)   \\
\hline
\underline{A}BA  & $\sqrt{11/3}\sim 1.91$ & $+2$ & 6 & 
(1,0,$2\sqrt{\frac{2}{3}}$)   \\
\hline
AB\underline{A}  & $\sqrt{11/3}\sim 1.91$ & $-2$ & 6 & 
(1,0,$-2\sqrt{\frac{2}{3}}$)   \\
\hline
All structures  & 2 & $0$ & 6 & 
(2,0,0)   \\
\hline
All structures  & $\sqrt{7}\sim2.64$ & $0$ & 12 & 
(2,$\sqrt{3}$,0)   \\
\hline
\end{tabular}
\caption{Beyond next and next-nearest neighbours. The `local structure' specifies the local layer pattern for which a particular neighbour appears. The reference layer is underlined. The other columns are as defined in Tab.~\ref{tab.neighbours1}.}
\label{tab.neighbours2}
\end{table}

\section{Tight binding setup}
In the tight binding approach described in the main text, we have implicitly assumed that orbitals on distinct sites are orthogonal. We now relax this condition and demonstrate that the gauge symmetry still survives. We follow a general tight binding prescription, e.g., as described in Ref.~\cite{Marder}. 
We begin with $s$ orbitals denoted as $\vert u, v, w,\alpha\rangle$, where $(u,v,w)$ specify the unit cell and $\alpha$ specifies a basis atom. We use $\hat{H}$ to denote the Hamiltonian for a single electron in a periodic potential induced by static nuclei. The tight binding description can be constructed using two types of coefficients that we denote as:
\begin{itemize}
\item Direct overlap: $d_{u,v,w,\alpha;a',v',w',\beta} = (-)\langle u,v,w,\alpha \vert u',v',w',\beta\rangle$. 
\item Hopping: $t_{u,v,w,\alpha;u',v',w',\beta} = (-)\langle u,v,w,\alpha \vert ~\hat{H} ~\vert u',v',w',\beta \rangle$.
\end{itemize}
The values of these coefficients are strongly constrained by geometry and symmetry. For example, due to the spherical symmetry of $s$ orbitals, ($d$,$t$) take the same values on all nearest neighbours. Below, we will only consider overlaps and hopping between 1nn and 2nn. This is justified when the atoms are well separated so that the local wavefunctions resemble atomic orbitals with exponential decay away from the nucleus.

As described in the main text, an ansatz for stationary states can be written as 
\begin{equation}
    \ket{\bm{k}}   \propto\sum_{u,v=1}^L\sum_{w=1}^{L_c}\sum_{\alpha=1}^M c_\alpha\,e^{i\bm{k}\cdot\bm{R}_{uvw\alpha}}\ket{u,v,w,\alpha},
\label{eq:bloch-ansatz}
\end{equation}
where ${\bm R}_{uvw\alpha}$ represents the position vector of the $\alpha^\mathrm{th}$ atom within the unit cell labelled by $(u,v,w)$. The coefficients $c_\alpha$ are to be determined such that this ansatz is a stationary state. This leads to a generalized eigenvalue problem at every point in the Brillouin zone, 
\begin{eqnarray}
H(\bm k) \{ c\} = E ~O(\bm k)~ \{c\},
\label{eq.geneig}
\end{eqnarray}
where $\{c\}$ is a vector of $c_\alpha$'s and $E$ is the energy eigenvalue. The matrix $H(\bm k)$ is the same as the hopping matrix defined in the main text. Its entries involve hopping amplitudes. 
The `overlap' matrix, $O(\bm k)$, contains entries that involve direct overlaps. Assuming that hopping and direct-overlap matrix elements are negligible beyond second nearest neighbours, the matrices have the form,
\begin{eqnarray}
      H(\bm k)=
    \begin{pmatrix}
    V_D & V_1 & 0 & \cdots & V_M^*\\
    V_1^* & V_D & V_2 & \ddots & 0\\
    0 & \ddots & \ddots & \ddots & 0\\
    \vdots & & V_{M-2}^* & V_D & V_{M-1} \\
    V_M & 0 & \cdots & V_{M-1}^* & V_D
    \end{pmatrix};~~~~~~~O(\bm k)=
    \begin{pmatrix}
    S_D & S_1 & 0 & \cdots & S_M^*\\
    S_1^* & S_D & S_2 & \ddots & 0\\
    0 & \ddots & \ddots & \ddots & 0\\
    \vdots & & S_{M-2}^* & S_D & S_{M-1} \\
    S_M & 0 & \cdots & S_{M-1}^* & S_D
    \end{pmatrix}.
    \label{eq.matrices}
\end{eqnarray}
This form is applicable when we have periodic boundaries along the stacking direction. In the case of open boundaries, the entries in the corners ($V_M$, $S_M$, $V_M^*$ and $S_M^*$) should be discarded. 

In each matrix, the diagonal entries are the same, given by 
\begin{eqnarray}
V_D = V_0-t\sum_{\bm{\eta}\in \mathcal{N}_\parallel}e^{i\bm{k}\cdot\vu{\eta}};~~~~~~~~~~~
S_D = S_0-d\sum_{\bm{\eta}\in \mathcal{N}_\parallel}e^{i\bm{k}\cdot\vu{\eta}}.
\end{eqnarray}
Here, $V_0 = \langle u,v,w,\alpha \vert ~\hat{H}~\vert u,v,w,\alpha\rangle$ and $S_0 = \langle u,v,w,\alpha \vert u,v,w,\alpha\rangle$ are on-site contributions. The remaining terms arise from overlaps with the six 1nn within the same layer. The parameters $(t,d)$ represent hopping and direct-overlap values between nearest neighbours.  

The off-diagonal entries depend on the chirality variables that describe the stacking sequence. In the hopping matrix, each off-diagonal entry takes one of two possible values:
\begin{eqnarray}
V_+=-t\sum_{\bm{\eta}\in \mathcal{N}_\perp^+}e^{i\bm{k}\cdot\vu{\eta}} -t'\sum_{\bm{\eta}\in \mathcal{N}_{2,\perp}^+}e^{i\bm{k}\cdot\vu{\eta}};~~~~~~~~~~~
V_-=-t\sum_{\bm{\eta}\in \mathcal{N}_\perp^-}e^{i\bm{k}\cdot\vu{\eta}}
 -t'\sum_{\bm{\eta}\in \mathcal{N}_{2,\perp}^-}e^{i\bm{k}\cdot\vu{\eta}}.
 \label{eq.V+-}
    \end{eqnarray}
Here, $t'$ represents the hopping amplitude between 2nn. Each term contains a summation over a set of three vectors, with $\mathcal{N}_\perp^\pm$ and $\mathcal{N}_{2,\perp}^\pm$ as defined in the main text. In the term proportional to $t$, the vectors connect to 1nn in the layer above. In the $t'$ term, we have vectors that connect to 2nn in the layer above. In both sets, each vector has the same the same $z$-component, $h \hat{z}$ with $h =\sqrt{\frac{2}{3}}$, corresponding to the perpendicular displacement between adjacent layers. We rewrite these summations as follows,
\begin{eqnarray}
V_+&=&-t e^{ik_z h}\sum_{\bm{\eta}\in 
\mathcal{N}_{xy}^+}e^{i\bm{k}\cdot\vu{\eta}} -t'e^{ik_z h}\sum_{\bm{\eta}\in \mathcal{N}_{2,{xy}}^+}e^{i\bm{k}\cdot\vu{\eta}} = e^{ik_z h} \times A_0 e^{i\phi};
\label{eq.Vplus}
\\
V_-&=&-te^{ik_z h}\sum_{\bm{\eta}\in \mathcal{N}_{xy}^-}e^{i\bm{k}\cdot\vu{\eta}}
 -t'e^{ik_z h}\sum_{\bm{\eta}\in \mathcal{N}_{2,{xy}}^-}e^{i\bm{k}\cdot\vu{\eta}}
 ~~=~~-te^{ik_z h}\sum_{\bm{\eta}\in \mathcal{N}_{xy}^+}e^{-i\bm{k}\cdot\vu{\eta}}
 -t'e^{ik_z h}\sum_{\bm{\eta}\in \mathcal{N}_{2,{xy}}^+}e^{-i\bm{k}\cdot\vu{\eta}}\nonumber \\
 &=&e^{ik_z h} \times A_0e^{-i\phi}
 .\label{eq.Vminus}
\end{eqnarray}
The vectors $\mathcal{N}_{xy}^+$, $\mathcal{N}_{xy}^-$, $\mathcal{N}_{2,{xy}}^+$ and $\mathcal{N}_{2,{xy}}^-$  represent the lateral components from the previously defined sets of vectors.
In the first line, after taking out the common factor of $e^{ik_z h}$, we are left with a complex number whose amplitude is denoted as $A_0$ and phase as $\phi$. In the second line, we have used the fact that vectors in $\mathcal{N}_{xy}^+$ are exactly opposite to those in $\mathcal{N}_{xy}^-$. Similarly, vectors in $\mathcal{N}_{2,{xy}}^+$ and $\mathcal{N}_{2,{xy}}^-$ are opposite to one another.
After extracting the common factor of $e^{ik_z h}$, we are left with the complex conjugate of $A_0 e^{i\phi}$. Comparing Eqs.~(\ref{eq.Vplus}-\ref{eq.Vminus}), we see that $V_+$ and $V_-$ are complex numbers with the same amplitude, but with different phases. This is true for any momentum $\bm k$. In the main text, we have asserted that 
$V_+ (\bm k) = V_-(\bm k)e^{2i\phi(\bm k)}$. 
By comparing Eqs.~\ref{eq.Vplus} and \ref{eq.Vminus}, we see that $\phi$ here is the same as $\phi(\bm k)$.

We now consider off-diagonal elements in the overlap matrix,
\begin{eqnarray}
S_+=-d\sum_{\bm{\eta}\in \mathcal{N}_\perp^+}e^{i\bm{k}\cdot\vu{\eta}} -d'\sum_{\bm{\eta}\in \mathcal{N}_{2,\perp}^+}e^{i\bm{k}\cdot\vu{\eta}};~~~~~~~~~~~
S_-=-d\sum_{\bm{\eta}\in \mathcal{N}_\perp^-}e^{i\bm{k}\cdot\vu{\eta}}
 -d'\sum_{\bm{\eta}\in \mathcal{N}_{2,\perp}^-}e^{i\bm{k}\cdot\vu{\eta}},
 \label{eq.S+-}
\end{eqnarray}
where $d$ and $d'$ represent the overlap integrals between nearest and next-nearest neighbours respectively. The summations are the same as those in Eq.~\ref{eq.V+-} above. Using the known forms of 1nn and 2nn vectors, we write
\begin{eqnarray}
 S_+&=&d~ e^{ik_z h}\sum_{\bm{\eta}\in \mathcal{N}_\parallel^+}e^{i\bm{k}\cdot\vu{\eta}} +d'~e^{ik_z h}\sum_{\bm{\eta}\in \mathcal{N}_{2,\parallel}^+}e^{i\bm{k}\cdot\vu{\eta}} = e^{ik_z h} ~\times~B_0 e^{i\xi};
\label{eq.Splus}\\
\nonumber S_-&=&-d~e^{ik_z h}\sum_{\bm{\eta}\in \mathcal{N}_\parallel^-}e^{i\bm{k}\cdot\vu{\eta}}
 -d'~e^{ik_z h}\sum_{\bm{\eta}\in \mathcal{N}_{2,\parallel}^-}e^{i\bm{k}\cdot\vu{\eta}}
 ~~=~~-d~e^{ik_z h}\sum_{\bm{\eta}\in \mathcal{N}_\parallel^+}e^{-i\bm{k}\cdot\vu{\eta}}
 -d'~e^{ik_z h}\sum_{\bm{\eta}\in \mathcal{N}_{2,\parallel}^+}e^{-i\bm{k}\cdot\vu{\eta}}\\
&=& e^{ik_z h} ~\times~B_0 e^{-i\xi}
 .\label{eq.Sminus}
\end{eqnarray}
As with $V_+$ and $V_-$, $S_+$ and $S_-$ are complex numbers with the same amplitude, but with different phases. We represent the phase of $S_+$, after extracting a factor of $e^{ik_z h}$, as $\xi$. The phase of $S_-$, after removing the same factor, is $(-\xi)$.

\subsection{Gauge symmetry upon including direct overlaps}

Gauge symmetry arises from the invariance of the generalized eigenvalue problem in Eq.~\ref{eq.geneig} under a unitary transformation. We require a unitary matrix $W$ such that
\begin{eqnarray}
W^\dagger ~H_\sigma(\bm k)~ W  = H_\chi(\bm k);~~~~~~
W^\dagger ~O_\sigma(\bm k)~ W  = O_\chi(\bm k),
\label{eq.transforms}
\end{eqnarray}
where $H_\sigma(\bm k)$ and $O_\sigma(\bm k)$ are the hopping and overlap matrices associated with a close-packed structure encoded as $(\sigma_1, \ldots,\sigma_M)$ in H\"agg code notation. The corresponding matrices for a different close-packed structure encoded as $(\chi_1, \ldots,\chi_M)$ are $H_\chi(\bm k)$ and $O_\chi(\bm k)$. As described in the main text, we take $W$ to be diagonal matrix, $W = \mathrm{diag}\{e^{i\theta_1},e^{i\theta_2},\ldots,e^{i\theta_M}\}$. The diagonal terms in $H(\bm k)$ and $O(\bm k)$ are easily seen to be unchanged under the unitary transformation. In order to satisfy Eq.~\ref{eq.transforms}, we require (i) the off-diagonal elements of $H_\sigma$ to transform into those of $H_\chi$ \textit{and} (ii) those $O_\sigma$ to transform into entries in $O_\chi$. As argued in the main text, to satisfy (i), $\theta$'s must satisfy the following condition,
\begin{eqnarray}
    (\theta_{j+1}-\theta_{j}) = (\chi_j-\sigma_j) \phi.
    \label{eq.zeta}
\end{eqnarray}
In order to satisfy (ii), the $\theta$'s must satisfy
\begin{eqnarray}
    (\theta_{j+1}-\theta_{j}) = ( \chi_j-\sigma_j)\xi.
    \label{eq.xi}
\end{eqnarray}

\noindent \underline{Restricting to nearest neighbour bonds}: 
We first restrict our attention to 1nn bonds alone. This amounts to setting $t' = d' = 0$ in Eqs.~\ref{eq.V+-} and \ref{eq.S+-} above. With this choice of parameters, an off-diagonal term in $H(\bm k)$ has the same phase as the corresponding term in $O(\bm k)$. That is, at any given $\bm k$, $\phi = \xi$ where $\phi$ and $\xi$ are the phases defined in Eqs.~\ref{eq.Vplus}-\ref{eq.Vminus} and Eqs.~\ref{eq.Splus}-\ref{eq.Sminus} respectively. As $\phi=\xi$, any choice of $\theta$'s that satisfies Eq.~\ref{eq.zeta} will also satisfy Eq.~\ref{eq.xi}. The same choice of $W$ will satisfy both conditions in Eq.~\ref{eq.transforms}. The gauge symmetry holds even as the overlap matrix is taken into account. \\

\noindent \underline{Including second nearest neighbours}: We now allow for $t'$ and $d'$ to have non-zero values. We note that they are independent parameters, although both arise from overlaps between orbitals on second neighbours. Examining Eqs.~\ref{eq.Vplus}-\ref{eq.Vminus} and Eqs.~\ref{eq.Splus}-\ref{eq.Sminus}, we see that $\phi$ and $\xi$ are now independent phases. In order to satisfy the two conditions in Eq.~\ref{eq.transforms}, we must have 
\begin{eqnarray}
     (\theta_{j+1}-\theta_{j}) = (\chi_j-\sigma_j)\phi =( \chi_j-\sigma_j)\xi. 
\end{eqnarray}
This cannot be satisfied in general when $\phi$ and $\xi$ are distinct. It follows that a unitary matrix $W$, with the form given above, cannot satisfy both relations in Eq.~\ref{eq.transforms}. As a result, generalized eigenvalue problems for distinct stacking sequences are not unitary related.

\noindent \underline{Including third and further neighbours}: With further neighbours, the structure of the hopping and overlap matrices changes. We no longer have tri-diagonal forms as in Eq.~\ref{eq.matrices}. The gauge transformation discussed thus far cannot relate distinct close-packed structures.

\section{Degeneracy with periodic boundaries}
\begin{figure}
    \centering
    
\includegraphics[width=4in]{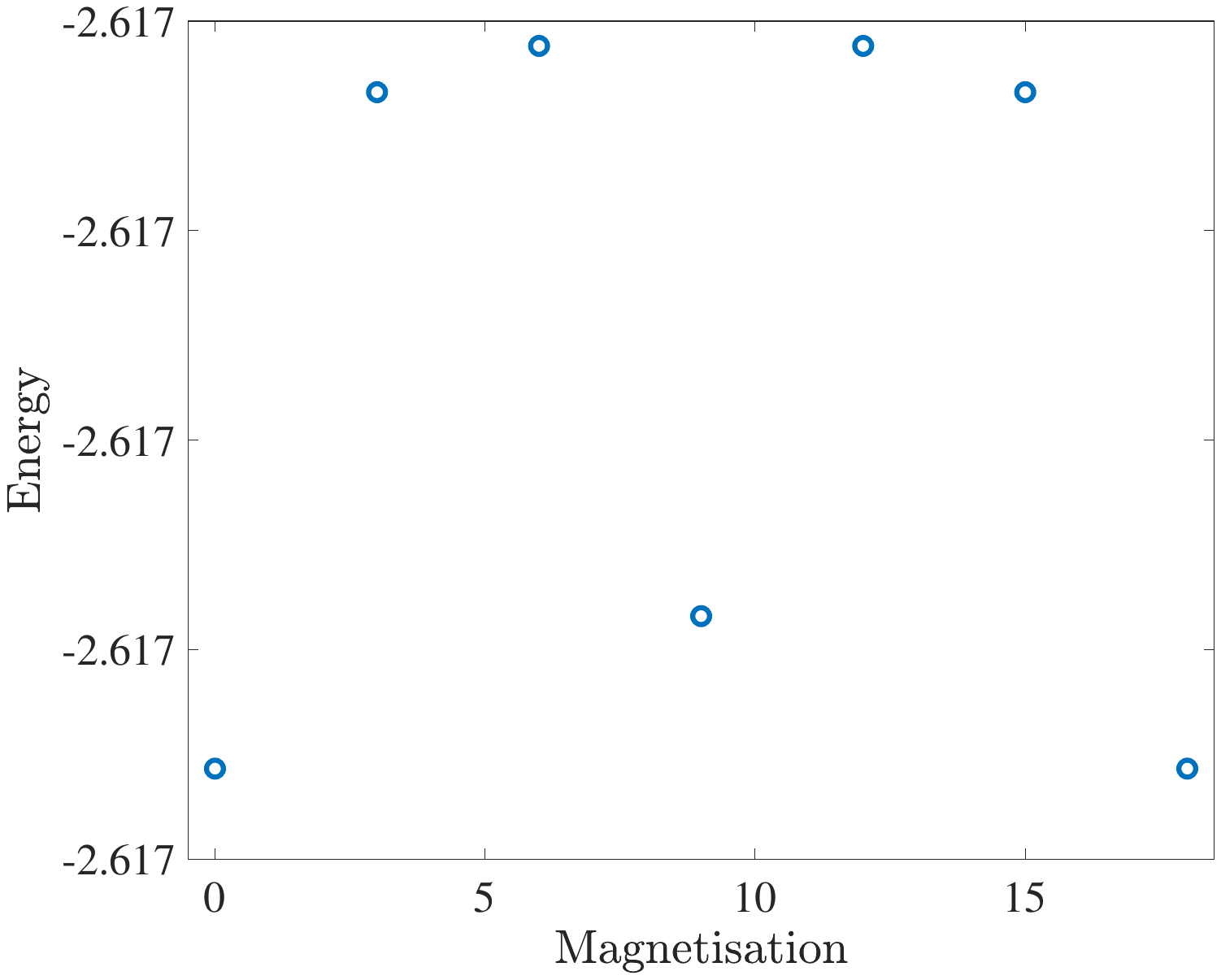}
    \caption{Energy as a function of net chirality for $M = 18$ sequences. The energies are calculated with $t$ set to unity, $t' = 0$. Overlaps between orbitals on distinct atoms have been neglected ($d=d'=0$). }
    \label{fig.EvsM}
\end{figure}

With $M$ layers and periodic boundaries along the stacking direction, close-packed structures are iso-energetic \textit{if} they have the same net chirality. As an example, we consider the problem of $M=18$ layers. The net chirality can then range from -18 to +18 in steps of 2. Fig.~\ref{fig.EvsM} plots the electronic energy as a function of the net chirality. The energy is calculated by filling all bands up to the Fermi level, with the Fermi level chosen so as to ensure a density of one electron per atom. As seen from the figure, the variation in energy is very small as net chirality is varied. 
The lowest energy states correspond to zero and extremal values (net chirality = $\pm$18).   

To explore implications for physical properties, we define `stacking degeneracy' as the number of close-packed arrangements that produce a certain net chirality. If the net chirality is $\Sigma$, the stacking degeneracy is ${}^MC_{(M+\Sigma)/2}$. We may define `stacking entropy' as the logarithm of this quantity. The lowest stacking entropy occurs for $\Sigma = \pm M$, where all chirality variables are the same. These states correspond to fcc order. In contrast, the largest entropy occurs for $\Sigma=0$ when half the chirality variables are +1 and the other half are -1. This set of states includes hcp order, occuring when chirality variables alternate as in a 1D Ising antiferromagnet.

The energy and stacking entropy are functions of net chirality. This suggests an interesting finite temperature phase diagram as free energy is a function of net chirality. With increasing temperature, we may expect multiple chirality sectors to appear at distinct critical temperatures. We emphasize that this scenario only applies when we have periodic boundaries along the stacking direction. This could be relevant to numerical studies, e.g., using molecular dynamics.

\section{Phonon band structures}

Our arguments regarding gauge symmetry and degeneracies are based solely on electronic energies. We now consider the phonon spectrum and its zero-point energy contribution. We use the \texttt{General Utility Lattice Program}~\cite{gale_MS_2003_GULP} to construct the dynamical matrix and to extract phonon eigenfrequencies. We assume that a pair of atoms located at $\bm{r},\bm{r}'$ interact through a central potential of the form
\begin{equation}
    \Phi_{\alpha\beta}(\bm{r},\bm{r}')
    =\sum_{\{\bm{\eta}\}}\left[(k_L-k_T)\,\frac{\bm{\eta}\bm{\eta}^T}{\eta^2}-k_T\,\delta_{\alpha\beta}\right]\delta_{\bm{r}+\bm{\eta},\bm{r}'},
    \label{eq:interatomic-potential}
\end{equation}
where $\alpha,\beta$ label Cartesian coordinates. The sum is over all nearest-neighbours of $\bm{r}$, with $\bm \eta$ running over all nearest-neighbour vectors. We neglect all couplings beyond 1nn; there is no interaction potential between second, third or farther neighbours. The longitudinal and transverse force constants, $k_L=0.382~\si{\electronvolt\per\angstrom\squared}$ and $k_T=-0.024~\si{\electronvolt\per\angstrom\squared}$, respectively, have been chosen to reproduce density functional theory results for the lattice constant and for phonon frequencies at $\Gamma,X,L$ in fcc Li~\cite{hutcheon_PRB_2019_fcc_Li}.

\begin{figure}[htbp]
    \centering
    \includegraphics[width=0.8\textwidth]{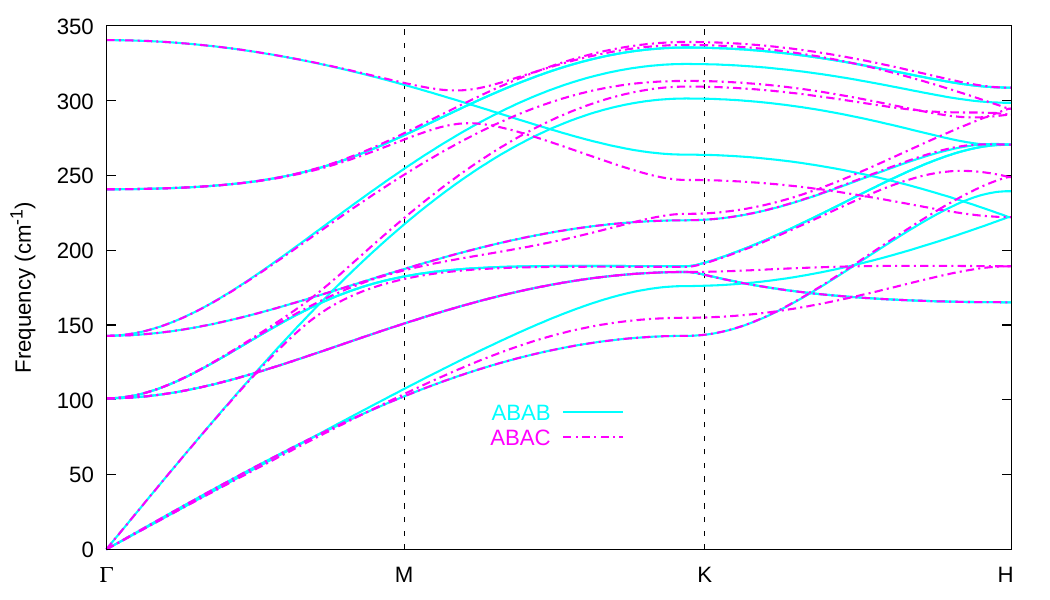}
    \caption{Phonon bandstructure of $ABAB$ and $ABAC$ assuming couplings to first nearest-neighbours only. The high-symmetry points refer to the hexagonal Brillouin zone, corresponding to the unit cell described in the main text.}
    \label{fig:phonon_bands}
\end{figure}

Fig.~\ref{fig:phonon_bands} shows the obtained phonon band structures for two $M=4$ stackings: $(ABAB)$ and $(ABAC)$. We have assumed periodic boundary conditions along the stacking direction. As seen from the figure, the phonon bands are \textit{not} degenerate. This is despite the fact that both stackings have the same net chirality. 
Mathematically, Eq.~\ref{eq:interatomic-potential} has the same structure as the Slater-Koster overlaps between $p$-orbitals. The gauge symmetry holds for $s$ orbitals, but not with $p$ orbitals. This difference emerges from the directional nature of overlaps in $p$ (and higher) orbitals. 

When the temperature is close to absolute zero, phonon bands are not occupied. However, they give rise to a zero-point energy contribution -- a quantum effect arising from zero-point vibrations of atoms within the solid. We evaluate zero point energy from the obtained phonon band structure. As an illustration, Tab.~\ref{tab:symmetry-breaking} lists the zero-point energies for lithium in three $M=6$ close-packed structures. The second column in the table lists the corresponding electronic energy, with contributions from all states below the Fermi level. We see that phonon zero point energies are a very small effect, amounting to a few percent of the electronic energy. The differences in zero point energy are even smaller, with magnitudes $\sim 0.001\%$ of the electronic energy. 

These values show that phonon zero-point energies are a small effect, their contribution to energy being much smaller than that of electrons. While gauge symmetry forces the electronic energy to be the same across close-packed structures, phonon zero-point energies can be different. These differences are very small, of the order of 0.05 meV. We may expect these differences to play a role when the temperature is below 5 K or so. They may select one particular close-packed structure over others.

\section{Higher orbitals and hopping to further neighbours}

\begin{figure}
    \centering
    \includegraphics[width=4in]{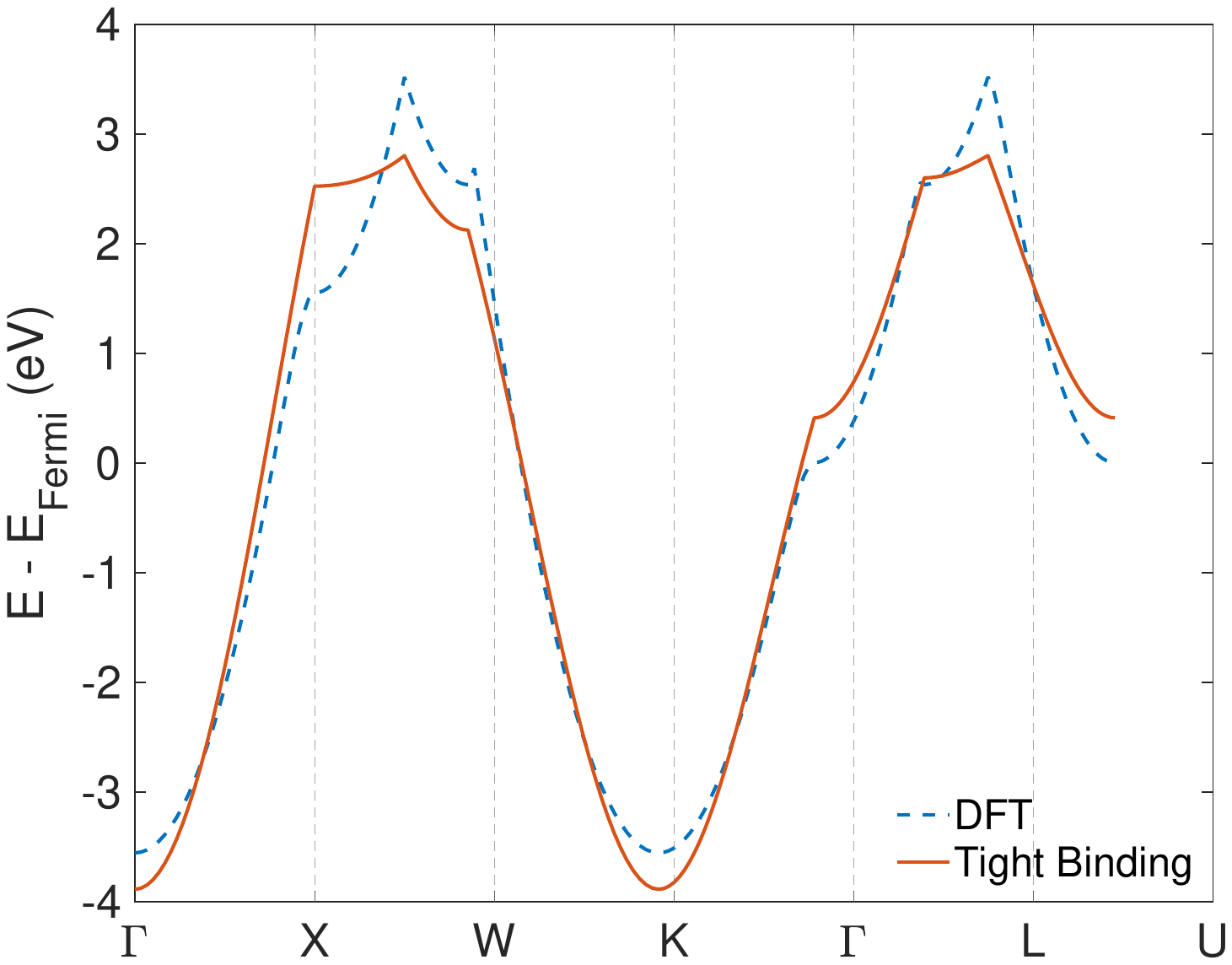}
    \caption{The fermi-energy-crossing band of fcc lithium. The band energies obtained from density functional theory (DFT) are compared to a tight binding fit. The latter involves a single $s$ orbital at each site, with hopping along 1nn and 2nn bonds. The best fit parameters correspond to $t=-0.53$ and $t'=0.07$. Direct overlaps are ignored. }
    \label{fig:TBDFT}
\end{figure}

In the arguments above, we have made two assumptions about the appropriate tight binding description of lithium and sodium. The first is the restriction to a single $s$ orbital per atom. This is consistent with well-known features of their band structure. For example, in the fcc structure, both lithium and sodium have a single band at the Fermi energy with a large $s$ orbital weight\cite{elatresh_PNAS_2017,Elatresh2020}. The second assumption is to only retain 1nn and 2nn bonds. The appropriateness of these two assumptions can be seen from Fig.~\ref{fig:TBDFT}. The figure shows the Fermi-energy-crossing band in fcc lithium obtained from density functional theory. This band is fit to the tight binding  result with one $s$ orbital and (1nn, 2nn) bonds, showing reasonable agreement. We now examine relaxing each of these two assumptions, focusing on consequences for the energetics of close-packing.

\begin{figure}
    \centering
    \includegraphics[width=6in]{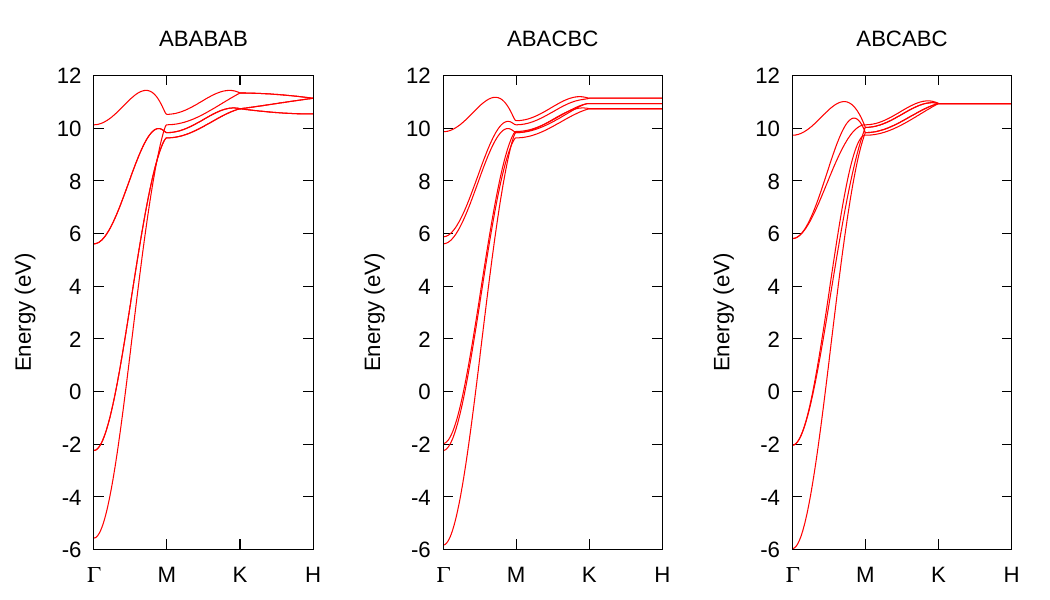}
    \caption{Effect of 3nn bonds: Band structure of select $M=6$ sequences with periodic boundaries. Direct overlaps are neglected. The two structures on the left have the same net chirality. All three band structures are very close, but with small differences among them. There is no symmetry-induced degeneracy.}
    \label{fig:3nn}
\end{figure}

Ref.~\onlinecite{papaconstantopoulos_Springer_2014} lists Slater Koster parameters for first, second and third neighbours in lithium, obtained by fitting to ab initio band structure in the bcc phase. Although we are interested in close-packed structures and not in bcc, we use these parameters as ballpark estimates for the true tight binding parameters. In Fig.~\ref{fig:3nn}, we plot the band structure obtained upon including 1nn, 2nn, 3nn bonds for three different close-packed structures. We retain a single $2s$ orbital per lithium atom. In the the tight binding Hamiltonian, we only keep one Slater Koster parameter corresponding to the $ss\sigma$ bond. The band structures are very close to another, but not identical. The band energies are particularly close between structures with the same net chirality. This demonstrates that 3nn hopping breaks the gauge symmetry weakly, leaving an approximate degeneracy. Put together, we conclude that 3nn (and further) bonds are a small effect that can induce weak degeneracy-breaking.

\begin{figure}
    \centering
    \includegraphics[width=6in]{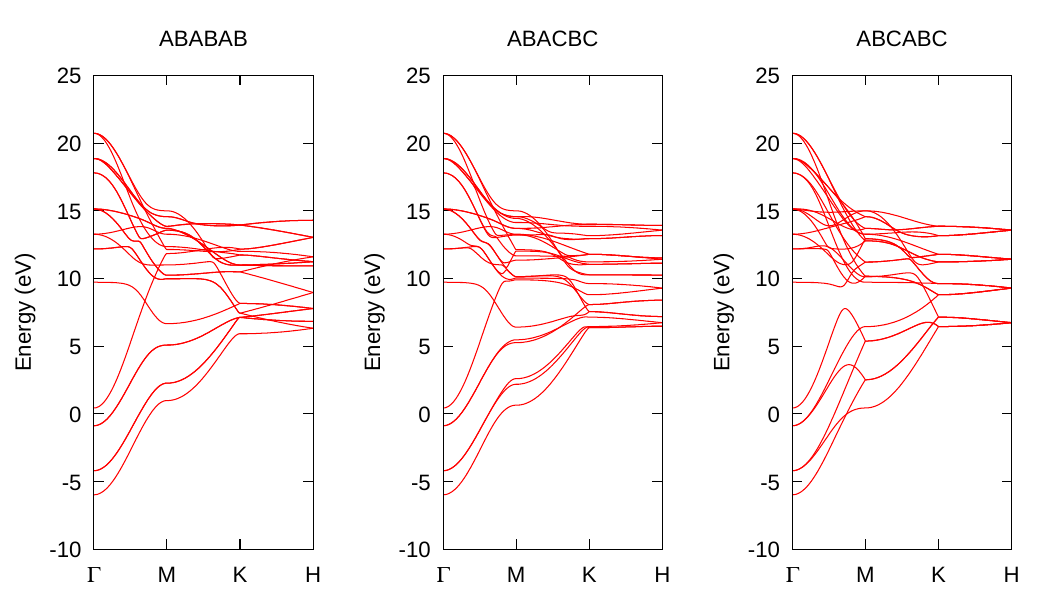}
    \caption{Effect of $p$ orbital admixture: Bandstructure of select $M=6$ sequences with periodic boundaries. Each atom contributes one $s$ and 3 $p$ orbitals. Direct overlaps are neglected. Hopping is truncated at the second nearest-neighbour. Slater-Koster parameters are chosen to reproduce known band structure of bcc lithium. 
    Inset: Bands over a narrow energy range, where discrepancies between $(ABABAB)$ and $(ABACBC)$ become visible.
    }
    \label{fig:sp3}
\end{figure}

We next consider the role of higher orbitals. For concreteness, we construct tight binding models with one $s$ and three $p$ orbitals at each atom. These can be viewed as the $2s$ and $2p$ valence orbitals of a lithium atom. We use Slater Koster parameters known from bcc lithium\cite{papaconstantopoulos_Springer_2014}, retaining $ss\sigma$, $sp\sigma$, $pp\sigma$ and $pp\pi$ parameters. Fig.~\ref{fig:sp3} shows the resulting band structure in three close-packed sequences of $M=6$ layers. Periodic boundary conditions are applied along the stacking direction. Two of the sequences (ABABAB and ABACBC) have the same net chirality, while the third differs (ABCABC). As seen in the figure, all three structures are very close in energy, but not identical. This is consistent with weak degeneracy breaking due to admixture between $s$ and $p$ orbitals.

We have discussed various effects that could break the gauge-symmetry-induced degeneracy: phonon zero-point energies, further neighbours and higher orbitals. Tab.~\ref{tab:symmetry-breaking} compiles estimates from these effects. The first column lists three $M=6$ sequences that serve as test cases. The second column lists the electronic energy from tight binding with a single $2s$ orbital, including 1nn and 2nn (no direct overlaps). This column reflects the degeneracy induced by gauge symmetry, with ABABAB and ABACBC having the same energy on account of having the same net chirality. The third column lists the electronic energy upon including $2s$ and $2p$ orbitals, with 1nn and 2nn bonds. The fourth column lists energies with a single $2s$ orbital, but including 3nn bonds. The fifth column lists energies with 1nn, 2nn and 3nn, including $2s$ and $2p$ orbitals. Finally, the last column lists the zero point energies of phonons. Columns 3-6 show small energy differences that arise from breaking of gauge symmetry.

\begin{table}
    \centering
    \begin{tabular}{|c|c|c|c|c|c|}
        \hline
        & 2nn, $s$ orbitals & 2nn, $sp^3$ orbitals & 3nn, $s$ orbitals & 3nn, $sp^3$ orbitals & Zero point energy \\
        \hline\hline
        $(ABABAB)$ & 41.277794 & 5.394210 & 41.002947 & 5.703231 & 0.241926 \\\hline 
        $(ABACBC)$ & 41.277794 & 5.284812 & 41.180439 & 5.387801 & 0.241895 \\\hline 
        $(ABCABC)$ & 41.271891 & 5.425520 & 41.271891 & 5.425520 & 0.241875 \\\hline 
    \end{tabular}
    \caption{Column~2: electronic ground state energies of select $M=6$ structures retaining only~$s$ orbital interactions and truncating hoppings at the second nearest-neighbour; \textit{i.e}. the assumptions under which there exists a gauge symmetry. Columns~3,4: electronic ground state energies after including~$p$ orbitals and third nearest-neighbours. Column~5: phonon zero-point energies. Electronic energies are calculated using a tight binding model with Slater-Koster parameters adapted to bcc~Li and a $20\times20\times20$ mesh of $k$-points over first Brillouin zone. All values in units of~eV.
    }
    \label{tab:symmetry-breaking}
\end{table}

\end{document}